\documentclass[
prc,twocolumn,superscriptaddress,showpacs,twoside,floatfix]{revtex4-1}

\usepackage{amsmath}
\usepackage{amssymb}
\usepackage{amsfonts}
\usepackage[dvips]{graphicx}
\usepackage{isotope}
\begin{document}

\title{High precision studies of soft dipole mode in two-neutron halo nuclei:
$^{6}$He case}

\author{L.V.~Grigorenko}
\email{lgrigorenko@yandex.ru}
\affiliation{Flerov Laboratory of Nuclear Reactions, JINR, 141980 Dubna, Russia}
\affiliation{National Research Nuclear University ``MEPhI'',
  115409 Moscow, Russia}
\affiliation{National Research Centre ``Kurchatov Institute'', Kurchatov
  sq.\ 1, 123182 Moscow, Russia}

\author{N.B.~Shulgina}
\affiliation{National Research Centre ``Kurchatov Institute'', Kurchatov
  sq.\ 1, 123182 Moscow, Russia}
\affiliation{Bogoliubov Laboratory of Theoretical Physics, JINR, 141980 Dubna,
Russia}

\author{M.V.~Zhukov}
\affiliation{Department of Physics, Chalmers University of Technology, 41296
  G\"{o}teborg, Sweden}

\date{\today.}

\begin{abstract}
The ``soft dipole'' E1 strength function is calculated for the transition from
the $^{6}$He $0^+$ ground state to the $1^-$ continuum $^{4}$He+$n$+$n$. The
calculations were performed within the hyperspherical harmonics formalism. The
sensitivity of the results to the $^{6}$He ground state structure and to final
state interactions, are analyzed. The large-basis calculations show the reliably
converged results for soft dipole strength function and for momentum
correlations of the $^{6}\mbox{He} \rightarrow \, ^{4}$He+$n$+$n$ dissociation
products. Transition mechanisms are analyzed based on the momentum correlations.
The comparison with experimental data is provided.
\end{abstract}

\maketitle


\section{Introduction}


The basic idea of the Soft Dipole Mode (SDM) is quite simple. Wave function (WF)
of weakly bound state has long asymptotic tail spreading in the classically
forbidden region (nucleon halo). Acting on such a WF by electromagnetic operator
(with a power dependence on radius) further enhances the asymptotic region and
we get a very long range ``source'', which populates the continuum. In this
situation the transition matrix element may get noticeable low-energy
enhancement even in the case of smooth (nonresonant) continuum in the final
state. For one-neutron haloes this scenario becomes important for binding
energies smaller than 1 MeV, providing the peak in the E1 strength function (SF)
at decay energies smaller than 1 MeV.

We would like to begin this work with some terminological note, which in reality
is deeply connected with the essence of the discussed problem. There exists
certain controversy about the idea of the SDM on which we would like to dwell a
little. Sometimes this phenomenon is characterized as ``soft dipole resonance''.
Such a notion contradicts the standard vision of ``resonance'' as an entity,
which is totally independent on the population mechanism. It stems, however,
from vision of the SDM as a ``low-energy offspring'' of the Giant Dipole
Resonance (GDR). The GDR phenomenon is not, strictly speaking, resonance itself,
but it unifies resonances of certain collective nature clustering in the certain
energy range. The two-body SDM evidently does not belong to this realm being a
single-particle ``geometric'' phenomenon. In contrast, the Pigmi Dipole
Resonance  (PDR) \cite{Paar:2007,Savran:2013} can be seen as a true collective
excitation connected with several neutrons contributing to formation of neutron
skin. In any case it should be understood that the existence of both the SDM and
the PDR excitations is based on the \emph{separation of scales} in the nucleon
WF. These are radial scales of halo nucleon (or skin nucleons) WF and radial
scale of the ``bulk'' of nucleons. The bulk of nucleons contributes to GDR
formation, while the nucleons of halo (or skin) produce the sizable low-energy
enhancement in the E1 strength function --- SDM (or PDR).

\begin{figure}
\begin{center}
\includegraphics[width=0.48\textwidth]{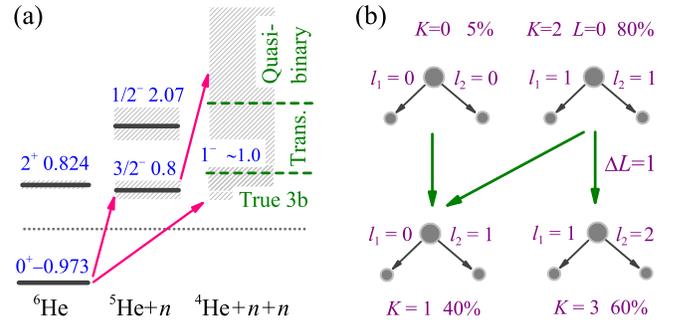}
\end{center}
\caption{(Color online) (a) Energy levels of $^{6}$He (left), $^{5}$He (middle),
and (right) population mechanisms for the soft dipole excitation in $^{6}$He at
different energies. (b) Population of the major configurations of the SDM source
(\ref{eq:sour-fun}) from the major configurations of the $^{6}$He g.s.\ WF
(\ref{eq:wf-bound}).}
\label{fig:scheme}
\end{figure}

In contrast to one-neutron halo case, the SDM in the case of three-cluster
systems (two-nucleon haloes) is quite complicated. (i) The SDM in the
three-cluster systems can not be attributed entirely to initial state geometry
as in the two-body case. The continuum dynamics in this case can not be
neglected if we would like to produce a decent approximation to the real
situation. (ii) The continuum dynamics in this case is an entangled mixture of
resonant and nonresonant dynamics. Qualitatively, in the odd-parity continuum
one of the nucleons populates a ``natural'' parity state where it has strong
resonating interaction with core ($l=1$ for $^{6}$He), while the other nucleon
is in nonresonating ``non-natural'' parity state (e.g.\ $l=0$, 2 for $^{6}$He),
see Fig.\ \ref{fig:scheme} (b) for illustration. Thus the three-body SDM can be
seen as a collective phenomenon  with only valence nucleons involved in the
collective motion. This form of continuum dynamics is especially difficult for
treatment and demands high calculation accuracy.

The soft dipole excitations of $^{6}$He were studied theoretically by different
methods
\cite{Danilin:1993,Cobis:1997,Danilin:1998,Myo:2001,deDiego:2010,Lay:2010,%
Descouvemont:2012,Myo:2014,Singh:2016}.
Shortcomings of these studies will be discussed later in present paper. We
obtain accurate fully converged results for E1 SF and well converged results
for three-body correlations. Reliable convergence allows to understand a lot of
problematic issues of the previous studies. An important aspect of the analysis
illustrated in Fig.\ \ref{fig:scheme} (a) is the transition from three-body
dynamics of SDM at low energies ($E_T \lesssim 1$ MeV) to semi-sequential
dynamics at high energies ($E_T \gtrsim 2.5$ MeV); three-body energy $E_T$ is
defined relative to the three-body breakup threshold.

The only experimental data available for the $^{6}\mbox{He} \rightarrow \,
^{4}$He+$n$+$n$ dissociation is the studies of Ref.\ \cite{Aumann:1999}. This
material was further elaborated in the review paper \cite{Aumann:2005}. The
three-body correlation aspect of these data was analyzed in Ref.\
\cite{Chulkov:2005}. The data are over 20 years old and quality of them are not
very high. It is also important to note that a lot of experimental efforts were
dedicated to the SDM in $^{11}$Li. The general experimental situation regarding
the Coulomb dissociation of $^{11}$Li is quite controversial \cite[and Refs.\
therein]{Nakamura:2006}. More recent experimental results on different inelastic
excitations of $^{11}$Li \cite{Kanungo:2015,Tanaka:2017} support the idea of
very pronounced SDM in $^{11}$Li but do not allow full quantitative description.
The $^{6}$He nucleus could have been a reference case for SDM studies in the
three-cluster systems, but very detailed and accurate experimental data are
needed, which are not available so far.

The interest to studies of SDM (or/and PDR) is partly based on the common
nowadays idea that the radiative capture rates for the three-body capture
processes can be based on the experimentally measured Coulomb dissociation cross
sections. The procedure can not be absolutely straightforward because it
involves an extrapolation from intermediate energies (available experimentally)
to quite low energies (contributing to the astrophysical capture rates at
temperatures of astrophysical interest). The prerequisite of such an
extrapolation is, of course, an accurate treatment of the E1 SF. Problems of
this treatment were discussed in the papers
\cite{Grigorenko:2006,Parfenova:2018} for the case of the $^{15}$O+$p$+$p
\,\rightarrow \, ^{17}$Ne+$\gamma$ process. A discussion of low-energy behavior
of the E1 SF for the case of the  $^{4}$He+$n$+$n \,\rightarrow \,
^{6}$He+$\gamma$ process will be given in the forthcoming paper
\cite{Grigorenko:2020b}.


\section{Theoretical model}


The formalism we apply here has already been used for studies of the soft dipole
excitation of $^{17}$Ne in Ref.\ \cite{Grigorenko:2006} and for the isovector
soft dipole excitation of $^{6}$Be in Ref.\ \cite{Fomichev:2012}. However, it
was given there briefly, so adding some more technical details is appropriate.
The hyperspherical harmonics (HH) method itself has already been described in
our previous works \cite{Danilin:1991,Grigorenko:2001,Grigorenko:2009c}, and the
detailes are provided here more for completeness of the description.


\subsection{Model for the $^{6}$He E1 dissociation process}


The bound $^{6}$He g.s.\ wave function is obtained in a $^{4}$He+$n$+$n$ model
by solving the homogeneous three-body Schr\"odinger equation (SE)
\begin{eqnarray}
\left[ \hat{H}_3 + V_3(\rho) + E_{\text{b}} \right] \Psi^{J_i M_i}_{\text{gs}} =
0 \,, \quad \nonumber \\
\hat{H}_3 = \hat{T}_3 + V_{cn_1}(\mathbf{r}_{cn_1}) +
V_{cn_2}(\mathbf{r}_{cn_2}) + V_{n_1n_2}(\mathbf{r}_{n_1n_2})\,,\quad
\label{eq:shred-gs}
\end{eqnarray}
see also papers \cite{Danilin:1991,Zhukov:1993,Zhukov:1993a}. The ideology of
our approach is that the three-body formalism theoretically
``extrapolates'' the properties of the two-body subsystems [these are introduced
via the phenomenologically defined pairwise potentials
$V_{ij}(\mathbf{r}_{ij})$] to the properties of the ``composite'' three-body
systems. This works well for systems with developed clusterization and strongly
bound clusters. $^{6}$He is one of the best systems appropriate for such
theoretical studies. Nevertheless, this description is never perfect and for
careful calculations of certain observables we need to fit the basic properties
of the three-body system (binding energy for the g.s.) to the experimental ones.
For this reason phenomenological three-body potential $V_3(\rho)$ depending on
the hyperradius only is added to the Schr\"odinger equation.

To obtain the E1 strength function we solve the following inhomogeneous SE
\begin{equation}
\left[ \hat{H}_3 + \tilde{V}_3(\rho)-E_T \right]  \Psi_{M_i m}^{JM(+)} =
\mathcal{O}_{\text{E1},m} \Psi^{J_{i}M_i}_{\mbox{\scriptsize gs}} \,.
\label{eq:shred-e1}
\end{equation}
The phenomenological three-body potential which is appropriate for the continuum
is expected to be different from that for the ground state and also somehow
smaller $\langle \tilde{V}_3 \rangle \lesssim \langle V_3 \rangle $.

Within the hyperspherical method the three-body Jacobi vectors
\[
\{ \mathbf{X},\mathbf{Y} \} = \{X,\Omega _{x},Y,\Omega _{y}\}\,,
\]
and corresponding  hyperspherical variables in coordinate space
\[
\{\rho,\Omega _{\rho }\} \;, \quad \Omega _{\rho }= \{\theta _{\rho },\Omega
_{x},\Omega _{y}\} \,,
\]
are defined as
\begin{eqnarray}
\mathbf{X} & = & \mathbf{r}_{1}-\mathbf{r}_{2}\;,\quad
\mathbf{Y} = \frac{A_1\mathbf{r}_{1}+A_2
\mathbf{r}_{2}}{A_1+A_2}-\mathbf{r}_{3}\;,
\label{eq:jacobi-1} \\
\rho ^{2} & = & \frac{A_1 A_2}{A_1+A_2} X^{2} + \frac{(A_1+A_2)A_3}{A} Y^{2}
\nonumber \\
& = & (A_1 A_2 r_{12}^{2}+ A_2 A_3 r_{23}^{2} + A_3 A_1 r_{31}^{2}) /A   \;,
\label{eq:jacobi-2} \\
\theta _{\rho } & = & \arctan \left[ \sqrt{\frac{A_1 A_2
A}{A_3(A_1+A_2)^2}}\, \frac{X}{Y}\right] \, ,
\label{eq:jacobi-3}
\end{eqnarray}
where $A = A_1+A_2+A_3$. The three-body Schr\"odinger equations for core+$N$+$N$
systems are solved in the so-called ``T'' Jacobi system (core is particle number
3). Jacobi vectors and hyperangle  $\theta _{\rho }$ in the other Jacobi systems
can be obtained by cyclic permutations of the cluster coordinates and mass
numbers. The
hyperradius $\rho$ is invariant under permutations, see Eq.\
(\ref{eq:jacobi-2}).

The E1 transition operator has the following definition and relation to the
dipole operator
\[
\mathcal{O}_{\text{E1},m}=e \, \sum_{i=1,3}
Z_{i}\, r_{i} \,Y_{1 m}(\hat{r}_{i})=\sqrt{\frac{3}{4 \pi}} D_{m} \,,
\]
where $\mathbf{D} = \textstyle \sum_{i=1,3} e Z_{i} \mathbf{r}_{i}$. Attention
should be paid on a misprint in the definition of this operator in the paper
\cite{Parfenova:2018}, which, however, did not affect  the results of this
paper. For two-neutron halo case of $^{6}$He the dipole operator acts on the
core particle only
\begin{eqnarray}
\mathcal{O}_{\text{E1},m}=e\, Z_{3}\, r_{3} \,Y_{1m}(\hat{r}_{3}) =
Z_{\text{eff}}\, \rho \, \cos(\theta_{\rho}) \,
Y_{1m}(\hat{y})
\,, \nonumber \\
Z_{\text{eff}}^{2}=\frac{e^{2}\, Z_{3}^{2}\, (A_{1}+A_{2})}
{A_{3}(A_{1}+A_{2}+A_{3})} = \frac{e^{2}}{3}  \,.
\label{eq:dipole-oper}
\end{eqnarray}
For two-proton case this is also true, but with effective core charge
\[
Z_{3} \rightarrow Z_{3}-A_{3} \,,
\]
because in the center-of-mass we have the relation $A_{1}\mathbf{r}_{1}+
A_{2}\mathbf{r}_{2} \equiv - A_{3}\mathbf{r}_{3}$.

The three-body continuum WF $\Psi_{ M_{i}m}^{JM(+)}$ and the initial bound
state WF  $\Psi^{J_{i}M_{i}}_{\mbox{\scriptsize gs}}$ are defined as
\begin{eqnarray}
\Psi_{ M_{i}m}^{JM(+)}=C_{J_{i}M_{i}1m}^{JM} \, \rho^{-5/2} \sum_{K \gamma}
\chi_{JK \gamma}^{(+)}(\varkappa \rho) \, \mathcal{J}_{K \gamma
}^{JM}(\Omega_{\rho}) \,,\quad
\label{eq:wf-cont} \\
\Psi^{J_{i}M_{i}}_{\mbox{\scriptsize gs}}=\rho^{-5/2} \sum_{JK_{i}\gamma_{i}}
\chi_{J_{i}K_{i}\gamma_{i}}(\rho)\, \mathcal{J}_{K_{i}\gamma_{i}}
^{J_{i}M_{i}}(\Omega_{\rho}) \, . \quad
\label{eq:wf-bound}
\end{eqnarray}
The functions $\mathcal{J}_{K \gamma }^{JM}(\Omega _{\rho })$ are hyperspherical
harmonics coupled with spin functions to total spin $J$. ``Multiindex'' $\gamma$
denotes the complete set of three-body quantum numbers except the principal
quantum number $K$: for spinless core cluster $\gamma=\{L,S,l_x,l_y\}$.

For these WFs the Schr\"odinger equation (\ref{eq:shred-e1}) is reduced to a set
of coupled inhomogenious differential equations
\begin{eqnarray}
\left[  \frac{d^{2}}{d\rho^{2}}-\frac{\mathcal{L}(\mathcal{L}+1)}{\rho^{2}}
-2M(E_T-V_{K \gamma,K \gamma}(\rho)) \right]  \chi_{J K \gamma}^{(+)}(\varkappa
\rho) \nonumber \\
= 2M \! \! \sum_{K'\gamma' \neq K\gamma} V_{K'\gamma',K \gamma }(\rho)
\chi_{J K'
\gamma'}^{(+)}(\varkappa \rho) + 2M \phi_{K \gamma}(\rho)  \,, \quad
\label{eq:se-syst}
\end{eqnarray}
where $M$ is ``scaling'' mass, taken in this work as average nucleon mass in 
$^{6}$He. The generalized angular momentum is defined by the principal 
hyperspherical
quantum number $K$ as
\[
\mathcal{L}= K+3/2 \,.
\]
The partial wave decomposition of the SDM source is given by
\begin{eqnarray}
\phi_{K\gamma}(\rho) = Z_{\text{eff}}   \sum \limits_{K_{i}\gamma_{i}}
\left \langle K \gamma \left | \cos(\theta_{\rho}) \right | K_{i}\gamma_{i}
\right \rangle \nonumber \\
\times \left \langle J \gamma \left\Vert Y_{1}(\hat{y}) \right \Vert
J_{i}\gamma_{i} \right \rangle \,\rho\,\chi_{J_{i} K_{i}\gamma_{i}}(\rho) \,.
\label{eq:sour-fun}
\end{eqnarray}
The hyperspherical and reduced angular matrix elements are
\begin{eqnarray}
\left \langle K \gamma \left | \cos(\theta) \right | K_{i} \gamma_{i} \right
\rangle \nonumber \\
=\int_{0}^{\pi/2}\! d \theta_{\rho} \psi_{K}^{l_{x}l_{y}}(\theta_{\rho})
\psi_{K_{i}}^{l_{x}^{i}l_{y}^{i}} (\theta_{\rho}) \sin^{2} (\theta_{\rho})
\cos^{3} (\theta_{\rho})   \,, \nonumber \\
\left \langle J \gamma \left \Vert Y_{1}(\hat{y}) \right \Vert J_{i}\gamma_{i}
\right \rangle
=\hat{l}_{y}^{i}\hat{l}_{x}\hat{L} ^{i} \hat{L} \hat{S} \hat{J}^{i}
\hat{1}^{3}\,\delta_{S_{x}^{i}S_{x}}
\nonumber \\
\times \, \left \{
\begin{array}
[c]{ccc}%
l_{x}^{i} & l_{y}^{i} & L^{i}\\
0 & 1 & 1\\
l_{x} & l_{y} & L
\end{array}
\right \}  \left \{
\begin{array}
[c]{ccc}%
L^{i} & S^{i} & J^{i}\\
1 & 0 & 1\\
L & S & J
\end{array}
\right\}  \frac{C_{l_{y}^{i}010}^{l_{y}0}}{\sqrt{4\pi}} \,, \nonumber
\end{eqnarray}
where we use the shortcut notation $\hat{m}=\sqrt{2m+1}$.


The asymptotic expression for the WF $\chi_{J_{f}K_{f}\gamma_{f}}^{(+)}
(\varkappa\rho)$ is
\[
\chi_{J K \gamma}^{(+)}(\varkappa \rho)
= A_{J K \gamma}\, \mathcal{H}_{\mathcal{L}}^{(+)} (\varkappa \rho) \, .
\]
Here $\mathcal{H}^{(\pm)}_{\mathcal{L}} = \mathcal{N}_{\mathcal{L}} \pm i
\mathcal{J}_{\mathcal{L}} $ are the Riccati-Bessel functions of half-integer
index $\mathcal{L}$, with the long-range asymptotics $\sim \exp(\pm i \varkappa
\rho)$, describing the in- and outgoing three-body spherical waves. The outgoing
flux through the hypersphere of a large radius is
\[
 j_J=\frac{\varkappa}{M} \sum _{K \gamma}
\,\left\vert A_{JK\gamma}\right\vert ^{2} = \sum _{K \gamma}
\,\sqrt{\frac{2E_T}{M}} \left\vert A_{JK\gamma}\right\vert ^{2} \, ,
\]
and the E1 strength function is expressed via this flux as
\begin{equation}
\frac{dB_{\text{E1}}}{dE_T} = \frac{1}{2 \pi} \sum_J \frac{2J+1}{2J_i+1} \, j_J
\, .
\label{eq:dbde-1}
\end{equation}

Let's also establish a connection with more ordinary formalism expressing
the E1 strength function in terms of the matrix elements of the dipole operator.
Within the Green's function formalism for coupled channel differential equations
the asymptotic coefficient can be expressed as
\[
A_{J K \gamma} = - \frac{2M}{\varkappa}\int d \rho \, \sum_{K'\gamma'}
\chi_{JK \gamma,K'\gamma'}(\varkappa \rho)\,  \phi_{K'\gamma'}(\rho)\, ,
\]
where $\chi_{JK\gamma,K^{\prime}\gamma^{\prime}}$ is solution of the
\textit{homogeneous} part of equations (\ref{eq:se-syst}) diagonalizing the $3
\rightarrow 3$ elastic scattering S-matrix
\begin{eqnarray}
S_{K \gamma,K' \gamma'} & = & \exp(2i \delta_{K \gamma,K' \gamma'}) \,,
\nonumber \\
\chi_{J K \gamma,K' \gamma'}(\varkappa \rho) & = &
\exp(i \delta_{K \gamma,K' \gamma'})[
\mathcal{J}_{\mathcal{L}'}(\varkappa \rho) \cos(\delta_{K \gamma,K'\gamma'})
\nonumber \\
& + & \mathcal{N}_{\mathcal{L}'}(\varkappa \rho) \sin(\delta_{K \gamma,
K'\gamma'})
 ] \,. \nonumber
\end{eqnarray}
Then with definitions
\begin{eqnarray}
A_{J K \gamma} & = & - \frac{2M}{\varkappa} \, \sqrt{\frac{\pi}{2}} \,
M_{J K \gamma} \,, \nonumber \\
M_{JK \gamma} & = &
\sum _{K'\gamma',K_{i}\gamma_{i}} \left \langle K' \gamma' \left \vert
\cos(\theta) \right \vert  K_{i} \gamma_{i} \right \rangle
\left \langle J \gamma' \left \Vert Y_{1}(\hat{y}) \right \Vert J_{i}\gamma_{i}
\right \rangle     \nonumber \\
& \times & \int d \rho \,\sqrt{\frac{2}{\pi}}\,
\chi_{JK\gamma,K'\gamma'}(\varkappa \rho) \, \rho
\,\chi_{J_{i}K_{i}\gamma_{i}}(\rho)  \,.
\nonumber
\end{eqnarray}
one gets the conventional expression for the E1 strength function
\[
\frac{dB_{\text{E1}}}{dE_T} = \sum_J \frac{2J+1}{2J_i+1} \sum_{K\gamma}
\sqrt{\frac{M}{2E_T}} \, \left \vert M_{J K \gamma} \right \vert ^{2} \,,
\]
which is equivalent to Eq.\ (\ref{eq:dbde-1}). However, the solution of
\emph{inhomogenious} set of equations (\ref{eq:se-syst}) is found to be
technically preferable.

It is easy to find out that energy integrated value of the E1 strength function
is connected with the ground state rms value of the core distance $\left \langle
r_{3}^{2} \right \rangle$ from the cms of the whole three-body system.
\[
\int \frac{dB_{\text{E1}}}{dE_T} \, dE_T = \frac{3}{4\pi} \,e^{2}\, Z_{3}^2 \,
\left \langle r_{3}^{2} \right \rangle \,.
\]
This is so-called non-energy-weighted (NEW) E1 sum rule, which can be used for
cross-check of the theoretical calculations as well as for determination of
the ground state geometry from experimental data.


\subsection{Momentum distributions}


To define momentum distributions of the three-body decay products we should
introduce  Jacobi vectors $\{\mathbf{k}_{x} ,\mathbf{k}_{y}
\}$ in the momentum space and hyperspherical variables $\{\varkappa,\Omega
_{\varkappa } \}$
\begin{eqnarray}
\mathbf{k}_{x} & = & \frac{A_2}{A_1+A_2} \mathbf{k}_{1}
- \frac{A_1}{A_1+A_2} \mathbf{k}_{2} \, , \nonumber \\
\mathbf{k}_{y} & = & \frac{A_3}{A}\left( \mathbf{k}_{1}
+\mathbf{k}_{2}\right) -\frac{A_1+A_2}{A}\mathbf{k}_{3} \, ,   \nonumber \\
\varkappa ^{2} & = & 2ME_T=2M(E_{x}+E_{y}) \nonumber \\
& = & \frac{A_1 + A_2}{A_1 A_2}\, k_{x}^{2}+\frac{A}{(A_1 + A_2)A_3}
\,k_{y}^{2}\,,  \\
\Omega _{\varkappa } & = & \{\theta _{\varkappa},\Omega _{k_{x}},\Omega
_{k_{y}}\}\;,\quad \theta _{\varkappa}=\text{arctan} \left [\sqrt{E_{x}/E_{y}}
\right ]\,. \quad
\label{eq:jac-mom}
\end{eqnarray}
For the fixed decay energy, the three-body correlations are defined by five
parameters of $\Omega _{\varkappa }$. It is more practical to split the
correlation space into ``internal'' correlations (relative motion of three
particles) and ``external'' correlations (orientation of the decay plane in the
space). It is convenient to describe internal correlations with two parameters
$\{ \varepsilon, \cos (\theta _{k}) \}$, where  $\varepsilon
$ is the energy distribution between $X$ and $Y$ subsystems and
$\theta _{k}$ is the angle between the Jacobi momenta:
\begin{equation}
\varepsilon =\frac{E_{x}}{E_{T}} \; ,\qquad \cos (\theta _{k})=\frac{(
\mathbf{k}_{x}, \mathbf{k}_{y} ) }{k_{x}\,k_{y}}\,.
\label{eq:corel-param}
\end{equation}
These parameters can be constructed in any of three Jacobi systems. The
correlations constructed in different Jacobi systems are just different
representations of the same physical picture. However, different aspects of the
correlations may be better revealed in a particular Jacobi system. For the
core+$N$+$N$ systems there are two non-equivalent Jacobi systems: ``T'' and
``Y'' (the correlation information for the second ``Y'' system is the same).

The external correlations are connected with spin alignment of three-body
systems populated in reactions. Practical significance of such studies for the
three-body systems is discussed in the papers
\cite{Golovkov:2004,Golovkov:2005,Sidorchuk:2012,Chudoba:2018} and in the review
\cite{Grigorenko:2016}. No information of this kind is available for the
electromagnetic dissociation (EMD) of three-body systems and no further
discussion of this topic will be provided here. However, we should emphasize
that the relevant theoretical methods are already well developed and have proven
to be useful in many experimental situations. So, the application of the
corresponding analysis to the prospective EMD dissociation data is encouraged.


\subsection{Potentials}


We follow potential prescription for $A=6$ systems which has shown to be
efficient in Refs.\
\cite{Danilin:1991,Zhukov:1993,Grigorenko:2009c,Ershov:2010}.

The $NN$ potential is taken either as a simple $s$-wave single-Gaussian form
BJ (from the book of Brown and Jackson \cite{Brown:1976})
\begin{equation}
V_{nn}(r)=V_{0}\exp (-r^{2}/r_{0}^{2})\;,
\label{eq:bj}
\end{equation}
with $V_{0}=-31$ MeV and $r_{0}=1.8$ fm, or the realistic ``soft-core''
potential GPT (Gogny-Pires-de Tourreil \cite{Gogny:1970}).

In the $\alpha $-$n$ channel we use an $\ell$-dependent potential SBB
(Sack-Biedenharn-Breit \cite{Sack:1954})
\begin{equation}
V_{\alpha n}(r)=V_{c}^{(\ell)}\exp
(-r^{2}/r_{0}^{2})+(\mathbf{\ell}\cdot\mathbf{s})\,
V_{\ell s}\exp(-r^{2}/r_{0}^{2})\;,
\end{equation}
where $r_{0}=2.30$~fm, $V_{c}^{(0)}=50$~MeV, $V_{c}^{(1)}=-47.32$~MeV,
$V_{c}^{(2)}=-23$~MeV, and $V_{ls}=-11.71$~MeV.

To get the phenomenological binding-energy correction for $^{6}$He g.s.\ an
additional short-range three-body potential $V_3$ in Eq.\ (\ref{eq:shred-gs}) is
used in the form
\begin{equation}
V_{3}(\rho )=\delta _{K\gamma ,K^{\prime }\gamma ^{\prime
}} \, V_{3}^{(0)}/[1+\exp ((\rho -\rho _{0})/d_{3})]\;,  \label{eq:pot3}
\end{equation}
where $\rho _{0}=2.5$~fm and $d_{3}=0.4$~fm. This ``short-range'' three-body
potential (note also the small diffuseness) does not distort the interactions in
the subbarrier region which was found to be important for consistent studies of
the asymptotic WF properties, see, e.g., the discussion in
Ref.~\cite{Grigorenko:2007}.

We do not have clear physical motivation for introducing $\tilde{V}_3$ in Eq.\
(\ref{eq:shred-e1}). However, arbitrary variation of this potential is used in
Sec.\ \ref{sec:what} for studies of characteristic sensitivities of the
theoretical model.


\subsection{$^{6}$He ground state wave function}

\begin{figure}
\begin{center}
\includegraphics[width=0.48\textwidth]{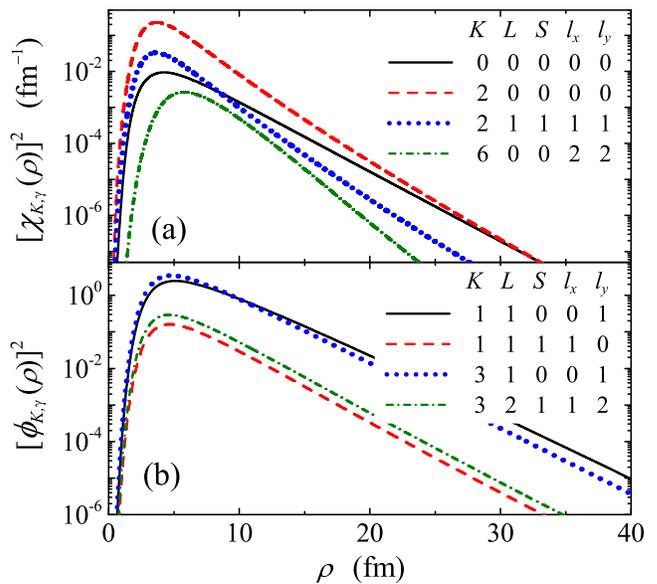}
\end{center}
\caption{Main components (squared) of the $^{6}$He g.s.\ WF (a) and the source
function Eq.\ (\ref{eq:sour-fun}) for the $1^-$ continuum (b).}
\label{fig:he6-wf-source}
\end{figure}

Different aspect of the $^{6}$He g.s.\ WF was studied in the hyperspherical
harmonics method several times \cite{Danilin:1991,Zhukov:1993}.
The obtained $^{6}$He and $^{6}$Li g.s.\ WF were tested against various
observables in several works
\cite{Danilin:1991,Danilin:1991a,Zhukov:1993a,Grigorenko:1998a,Danilin:1998,%
Ershov:2010}. They are known to provide consistent description of various
``long-range'' observables for the $^{6}$He and $^{6}$Li nuclei. The detailed
account of the isobaric symmetry of $^{6}$He and $^{6}$Be g.s.\ can be found in
the paper \cite{Grigorenko:2009c}. For that reason we give here the most
basic information about $^{6}$He g.s.\ and properties of the source function
induced by the dipole operator, see Fig.\ \ref{fig:he6-wf-source} and Table
\ref{tab:he6-wf-source}. In Sec. \ref{sec:what} the impact of the $^{6}$He g.s.\
WF variation on the E1 SF is studied. In this section the additional information
about $^{6}$He g.s.\ WF can be found, see Table \ref{tab:he6-wf-all}.


\subsection{Comment on Pauli principle treatment}


The three-body description of the six-body dynamics is an approximation, used by 
many scientific groups all over the world for $A=6$ systems. In our approach 
Pauli principle between ``valence'' neutrons and neutrons of the $\alpha$-core 
cluster is accounted approximately. The repulsive interaction is employed in the 
$s$-wave $\alpha$-$n$ channel, which well reproduces the experimental 
$\alpha$-$n$ scattering phases and largely prevent valence neutrons from 
entering core interior. Various ways of Pauli principle treatment both 
approximate and exact were used in the last three decades for studies of the 
$A=6$ systems. Different approaches could be more successfull for some aspects 
of dymamics and less to the others, but no ``silver bullet'' observable was 
found, which can confidently rule out some approaches.

The approximation used in this work is pragmatically justified by the mentioned 
above proper descriptions of various observables for $^6$He ground state. Our 
confidence in the three-body model applicability to E1 excitation in $^6$He is 
strongly supported by successful studies of continuum states in $^6$Be (both 
resonant and nonresonant) in Refs.\ 
\cite{Grigorenko:2009c,Fomichev:2012,Egorova:2012,Chudoba:2018}.
One may see in these works that even such subtle observables as very fine detals 
of three-body correlation patterns are nicely reproduced in the three-body model 
in spite of some deficiency in the Pauli principle treatment.
It should be also noted that E1 excitation is very peripheral process, becoming 
even more peripheral in the low-energy limit, where as we find in this work 
major computational problems take place. This makes the antisimmetrization issue 
presumably not of a prime importance for the problem we study.

\begin{table}[b]
\caption{Major components of the $^{6}$He g.s.\ WF (left 3 columns) and major 
components of the source function induced by the dipole operator (right 3 
columns). Relative probabilities $W$ are in percent and rms hyperradii
$\langle \rho \rangle$ are in fm.}
\begin{ruledtabular}
\begin{tabular}[c]{cccccc}
$K, L, S, l_x, l_y$ & $W $ &  $\langle \rho \rangle$ & $K, L, S, l_x, l_y$ & $W 
$ & $\langle \rho \rangle$ \\
\hline
 0   0   0   0   0 & 4.61 & 1.35  &  1  1  0  0  1  & 39.02 & 0.526 \\
 2   0   0   0   0 & 80.8 & 4.49  &  1  1  1  1  0  &  2.09 & 0.028 \\
 2   1   1   1   1 & 11.3 & 1.65  &  3  1  0  0  1  & 48.22 & 0.650 \\
 4   0   0   2   2 & 0.50 & 0.38  &  3  1  1  1  2  & 1.26  & 0.017 \\
 6   0   0   2   2 & 1.17 & 0.75  &  3  2  1  1  2  & 3.78  & 0.051 \\
 6   1   1   3   3 & 0.53 & 0.51  &  5  1  0  2  1  & 0.77  & 0.011 \\
%
%
\end{tabular}
\end{ruledtabular}
\label{tab:he6-wf-source}
\end{table}


\section{Convergence of SDM strength function}


The value $K_{\max}$ truncates the hyperspherical expansion in the system Eq.\
(\ref{eq:se-syst}). For each $K$ value all the possible basis states, namely all
the possible combinations of $l_x+l_y \le K$, are included in the HH expansion.

\begin{figure}
\begin{center}
\includegraphics[width=0.48\textwidth]{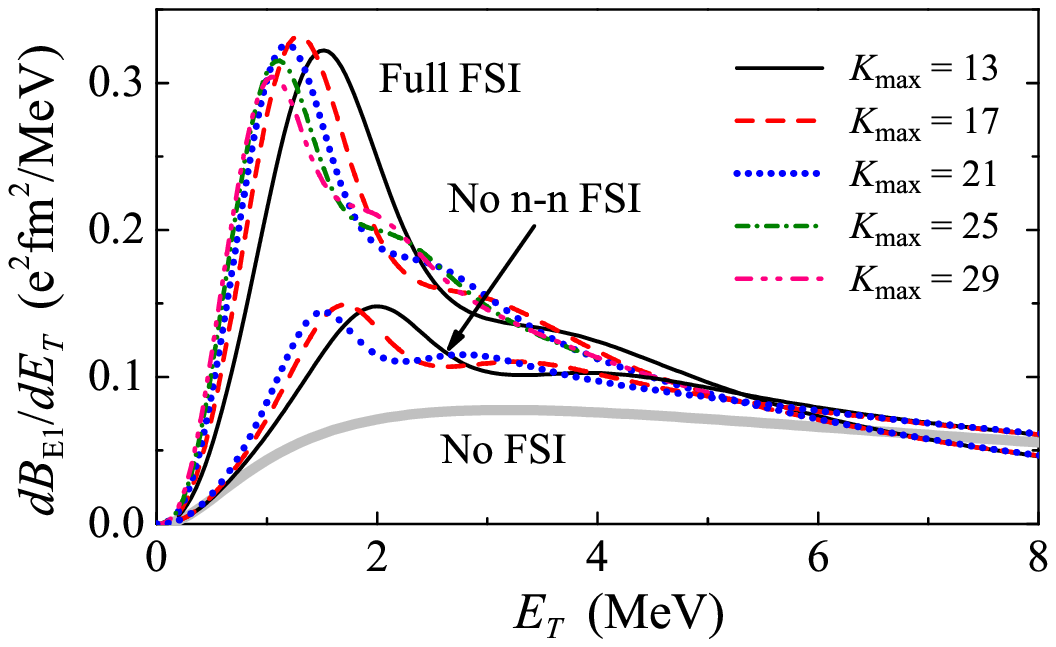}
\end{center}
\caption{Convergence of the E1 strength function calculations with and
without $n$-$n$ FSI for
$K_{\max}<30$.}
\label{fig:conv-low-k}
\end{figure}

The SE are solved up to $\rho_{\max}=400$ fm. The matching of the momentum
distribution is performed at $\rho_{\max}=70-90$ fm. At larger distances the
artefacts of the boundary conditions begin to arise. At $K_{\max}=25$ the number
of channels reaches 260, making further direct basis increase problematic. The
basis size can be effectively increased using the adiabatic procedure based on
the so-called Feshbach reduction (FR) \cite{Grigorenko:2007,Grigorenko:2009c}.
Feshbach reduction eliminates from the
total WF $\Psi =\Psi _{p}+\Psi _{q\text{,}}$ an arbitrary subspace $q$ using the
Green's function of this subspace:
\begin{equation}
H_{p}=T_{p}+V_{p}-V_{pq}G_{q}V_{pq}\;.
\end{equation}
In an adiabatic approximation, we can assume that the kinetic energy term is
small compared to the centrifugal barrier in the channels where this barrier is
large (these are evidently the channels with large $K$ values) and can be
approximated by a constant (``Feshbach energy'' $E_{f}$). In this approximation
the Green's function for the $q$ subspace can be defined by matrix inversion
from
\begin{multline}
(H-E_T)_{K\gamma ,K^{\prime}\gamma ^{\prime }} =
\left[ E_{f}-E_T+\frac{\mathcal{L}(\mathcal{L}+1)}{2M\rho ^{2}}\right]
\delta _{K \gamma,K' \gamma'} \\
+ V_{K \gamma ,K' \gamma '} = G_{K \gamma ,K^{\prime }\gamma ^{\prime }}^{-1}
\,.
\end{multline}
In this way the FR procedure is reduced to the construction of effective
three-body interactions
\begin{multline}
V_{K\gamma ,K^{\prime }\gamma ^{\prime }}^{\text{eff}} = V_{K\gamma ,K^{\prime
}\gamma ^{\prime }} \\
-\sum_{\bar{K}\bar{\gamma},\bar{K}^{\prime }\bar{\gamma}^{\prime }} V_{K \gamma
,\bar{K}\bar{\gamma}}G_{\bar{K}\bar{\gamma},\bar{K}^{\prime }
\bar{\gamma}^{\prime }} V_{\bar{K} ^{\prime }\bar{\gamma}^{\prime },K^{\prime
}\gamma ^{\prime }}\;.
\end{multline}
Summations over indexes with the bar are carried out for the eliminated channels
(the $q$ subspace). Technically, we eliminate all the channels with $K>K_{FR}$,
and the $K_{FR}$ value defines the sector of the HH basis where the
calculations remains fully dynamical. We take $E_{f}\equiv E_T$ in our
calculations as no significant sensitivity to this parameter in a broad
variation range was found.

\begin{figure}
\begin{center}
\includegraphics[width=0.48\textwidth]{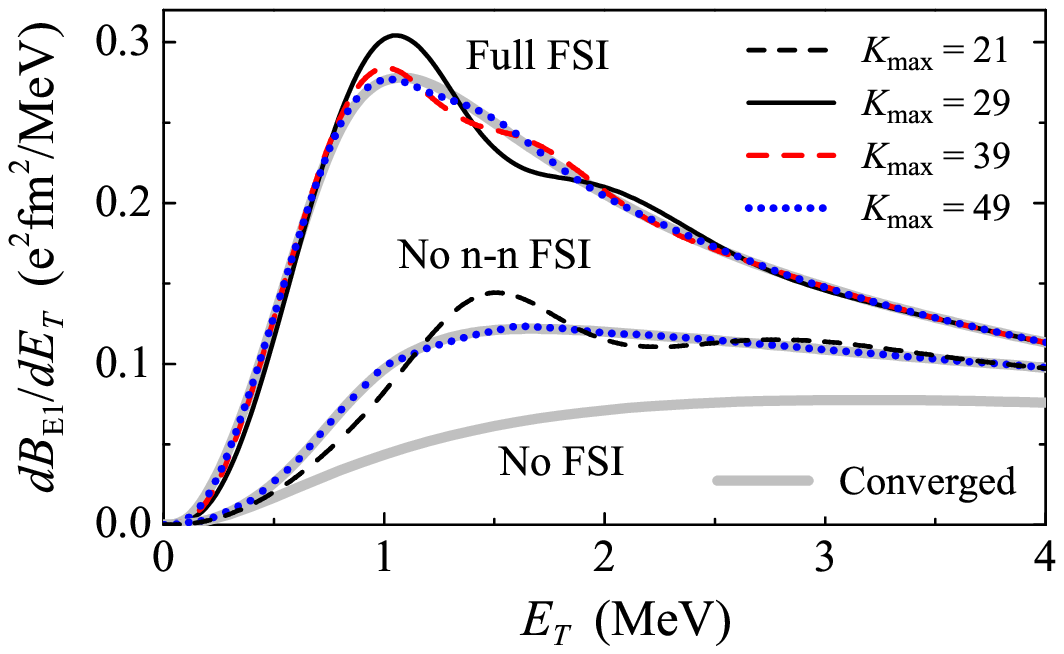}
\end{center}
\caption{Convergence of the E1 strength function calculations with and
without $n$-$n$ FSI for $K_{\max}>30$.}
\label{fig:conv-high-k}
\end{figure}

There are two forms of convergence to be studied to control the reliability of
the FR procedure. (i) One can gradually reduce $K_{\max }$ value for the fixed
$K_{FR}$ value taken as maximum attainable in the dynamic calculations. (ii) For
the maximum achieved $K_{\max }$ value one can gradually reduce $K_{FR}$ value
(using smaller and smaller dynamic basis size). The calculations of the strength
function for wide $E_T$ energy range is defined entirely by the $K_{\max}$
value. The basis size for the dynamical calculations can be taken as very modest
$K_{FR}=13-15$ without deterioration of the SF quality. However, the majority of
the presented calculations were performed with $K_{FR}=25$ which is very
reliable value. The largest  basis size is really needed (for $E_T > 0.4 $ MeV)
for calculations of the correlation patterns in the $\alpha$+$n$+$n$ continuum
since  the convergence for the correlations appears to be essentially slower
than for the strength function. These aspects of the convergence is discussed in
Section \ref{sec:mom-dis-conv}.

The convergence trends for small and large basis sizes are illustrated in Figs.\
\ref{fig:conv-low-k} and \ref{fig:conv-high-k}. We may see the following
important trends in the convergence patterns.

\noindent (i) The easiest task to get converged calculations is to remove FSI.
The convergence here is defined  by convergence of the source function expansion
which is practically achieved at $K_{\max}=5-7$, see Table
\ref{tab:he6-wf-source}.

\noindent (ii) The convergence of test calculations with $n$-$n$ FSI switched
off is achieved at $K_{\max}=30-40$. The conditions for calculations with full
three-body FSI is much worse, and complete convergence is achieved at
$K_{\max}=60-70$. The maximum basis size used in the calculations $K_{\max}=101$
which is very safe value for the SF calculations at $E_T > 0.4 $ MeV.

\noindent (iii) There is some form of systematic wavy behavior of the strength
function in the process of convergence. It seems that for small basis sizes the
SF oscillates around the final converged value. When the basis is increased,
these oscillations are shifted toward smaller energies and the magnitude of the
oscillations decreases. This form of the wavy behavior can be connected with
some kind of internal reflections in the system of three particles which arise
as an artefact of boundary conditions treatment in the situation of the basis
truncation.

\noindent (iv) Analogous picture of convergence spoiling the low-energy part of
the E1 SF was observed in the three-body calculations of SDM in $^{17}$Ne
\cite{Grigorenko:2006}. The cure for this problem was found in use of a model
with simplified Hamiltonian (without $p$-$p$ FSI ) which allowed to use the
exact three-body Green's function providing precise SF treatment
\cite{Grigorenko:2006,Parfenova:2018}. Application of such a model is based on
the fact that $p$-$p$ FSI was found to be not important for the low-energy E1 SF
calculations. In $^{6}$He this option is evidently not available, since there is
very large difference between calculations with and without $n$-$n$ FSI.

\noindent (v) Although the wavy behavior can be seen in
Fig.\ \ref{fig:conv-low-k} the position of peak in SF and the behavior of
the SF low-energy slope visually stabilize at $K_{\max}\sim 30$. For the basis
sizes achieved, these artificial waves are shifted to $E_T < 0.4 $ MeV.
The aspect of the low-energy SF convergence is quite intriguing and
discussed separately in Ref.\ \cite{Grigorenko:2020b}.

\begin{figure}
\begin{center}
\includegraphics[width=0.48\textwidth]{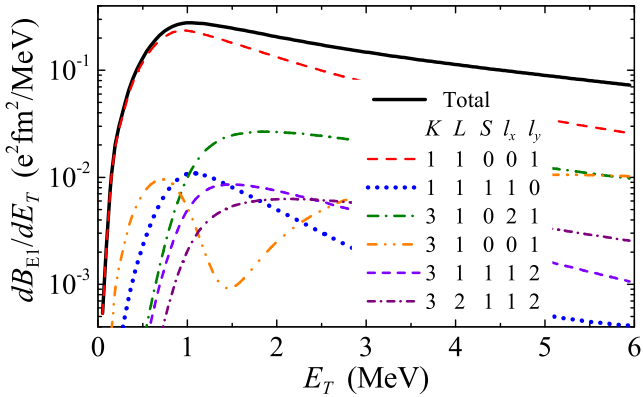}
\end{center}
\caption{The dominating contributions to the E1 strength function.}
\label{fig:sf-decomp}
\end{figure}

Comparison of calculations performed with quasirealistic GPT $n$-$n$ potential
and with simple central BJ $n$-$n$ potential provides very close results. This
is clear consequence of an extreme peripheral character of the SDM dynamics.
However, the convergence of calculations with GPT potential is much slower. For
that reason the largest-basis calculations were performed with the central
$n$-$n$ potential only.


\section{SDM decay dynamics}


\begin{figure*}
\begin{center}
\includegraphics[width=1.00\textwidth]{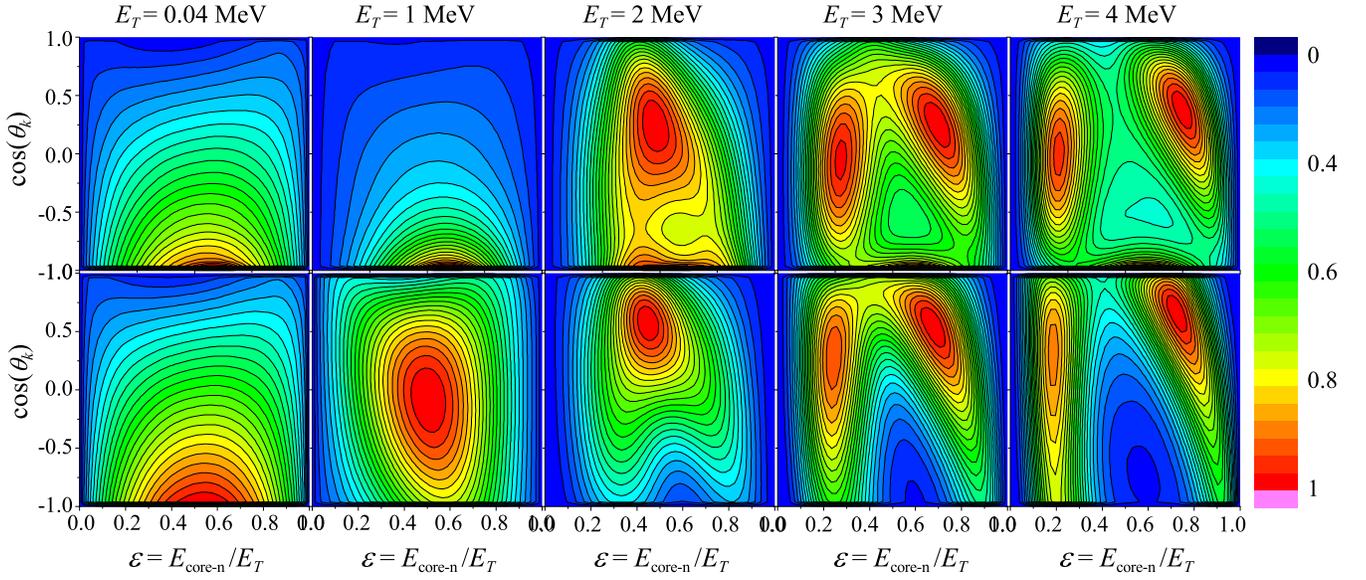}
\end{center}
\caption{Complete energy-angular three-body correlations for $^{6}$He SDM
transitions in the ``Y'' (right column) Jacobi systems. Upper row corresponds to
the full three-body calculation; bottom row corresponds to calculation without
$n$-$n$ FSI.
The columns correspond to decay energies  $E_T$ equal to 0.04, 1, 2, 3, and 4
MeV.}
\label{fig:en-ang-distr}
\end{figure*}

The partial wave decomposition of the $^{6}$He SDM SF is shown in Fig.\
\ref{fig:sf-decomp}. The low-energy part of the SF below $E_T=1$ MeV is strongly
dominated by the lowest hyperspherical component $\{K,\gamma \} =
\{1\,1\,0\,0\,1\}$. Sometimes this fact is interpreted as an opportunity to use
only one channel  (lowest possible channel with K = 1) in calculations of SDM.
However, this is not the case: although the relative weights of higher-$K$
channels are small, their cumulative effects to a large extent determine  the
``magnitude'' of the $K=1$ component in the low-energy domain.

The decay dynamics of the soft dipole mode can be clarified by momentum
distribution analysis of the decay products. The energy evolution of the
complete (energy-angular) three-body correlation patterns for $^{6}$He SDM is
illustrated in Fig.\ \ref{fig:en-ang-distr} for different decay energies. The
inclusive energy distributions are shown in Figs.\ \ref{fig:eps-t-low},
\ref{fig:eps-y-low}, \ref{fig:eps-y-high}, and \ref{fig:eps-t-high}. It can be
found that correlation patterns are
qualitatively different in three regions: (i) $E_T \lesssim 1$ MeV, (ii) $1
\lesssim E_T \lesssim 2.5$ MeV, (iii) $2.5 \lesssim E_T $ MeV.


\subsection{True three-body decay dynamics}


\begin{figure}
\begin{center}
\includegraphics[width=0.46\textwidth]{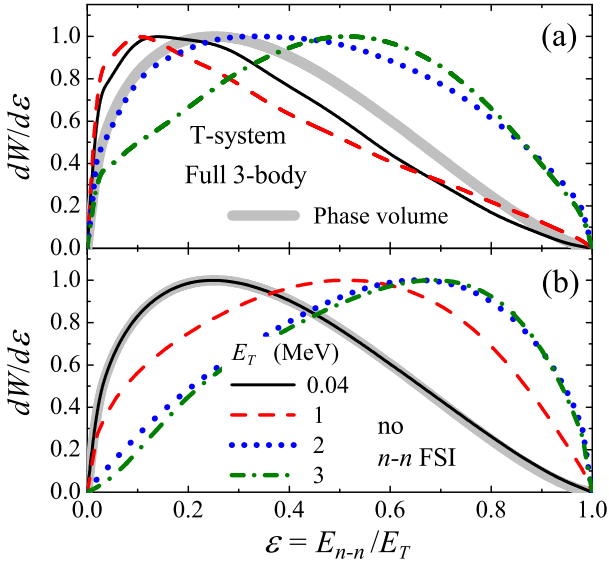}
\end{center}
\caption{Evolution of energy correlations over $E_T$ for $^{6}$He SDM
transitions in the ``T'' Jacobi system. The panels (a) and (b) corresponds to
full three-body calculation and ``no $n$-$n$ FSI'' case, respectively. The thick
gray curves show the phase volume  Eq.\ (\ref{eq:pv-2}).}
\label{fig:eps-t-low}
\end{figure}

The region (i) with $E_T \lesssim 1$ MeV corresponds to so-called true
three-body emission. This is a situation of essentially collective three-body
motion. Technically, it is expected that such a motion is well described by a
small number of HH terms. In the low-energy limit it should be just one term
with $K=K_{\min}=l_x(\min)+l_y(\min)$, most likely, the lowest hyperspherical
term (or, possibly, terms for $K_{\min}>0$).  The corresponding correlation
pattern is called ``three-body phase volume'' and it has meaning of phase volume
corrected for angular momenta. For single HH with definite $l_x$ and $l_y$
values the three-body phase space is
\begin{equation}
dW/d \varepsilon \sim \sqrt{\varepsilon^{1+2l_x}(1-\varepsilon)^{1+2l_y}} \,.
\label{eq:pv-1}
\end{equation}
It can be seen in Fig.\ \ref{fig:eps-t-low} that in the low-energy limit the
energy distribution in the Jacobi ``T'' system tends to phase volume for $[sp]$
configuration with $l_x=0$ and $l_y=1$
\begin{equation}
dW/d \varepsilon \sim \sqrt{\varepsilon(1-\varepsilon)^{3}} \,.
\label{eq:pv-2}
\end{equation}
For the ``no n-n FSI'' calculations this is exactly true for $E_T \lesssim 0.3$
MeV. In full three-body case there is a strong enhancement of the low-energy
part of the distribution due to the ``dineutron'' FSI. This effect is important
even at energy as low as $E_T \sim 0.04$ MeV and only for $E_T \lesssim 5$ keV
the three-body phase volume behavior is retained.

\begin{figure}
\begin{center}
\includegraphics[width=0.48\textwidth]{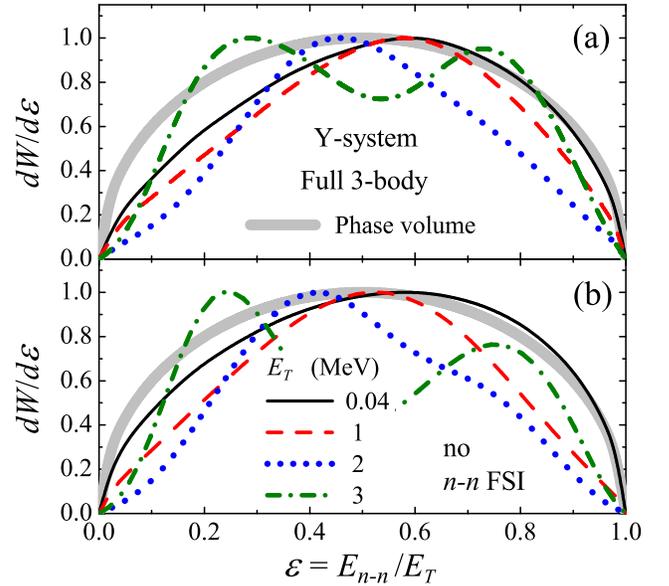}
\end{center}
\caption{Evolution of energy correlations over $E_T$ for $^{6}$He SDM
transitions in the ``Y'' Jacobi system. The panels (a) and (b) corresponds to
full three-body calculation and ``no $n$-$n$ FSI'' case, respectively. The thick
gray curves show the phase volume  Eq.\ (\ref{eq:pv-3}).}
\label{fig:eps-y-low}
\end{figure}

In a ``Y'' Jacobi system the corresponding three-body phase volume behavior is
defined by symmetry property of identical valence nucleons which leads to
$[sp]+[ps]$ configuration mixing with almost equal weights
\begin{equation}
dW/d \varepsilon \sim \sqrt{\varepsilon^{3}(1-\varepsilon)} +
\sqrt{\varepsilon(1-\varepsilon)^{3}} = \sqrt{\varepsilon(1-\varepsilon)} \,.
\label{eq:pv-3}
\end{equation}
This expression is equivalent to the most trivial three-body $s$-wave phase
volume ($l_x=0$ and $l_y=0$ case). It can be found in Fig.\ \ref{fig:eps-y-low}
that this simplistic expectation is well justified for $E_T \lesssim 0.3$ MeV.
For energies $E_T \lesssim 1$ MeV and above the energy distribution in the ``Y''
Jacobi system tends to relatively symmetric profiles with maximum at 
$\varepsilon \sim 0.5$, which are typical for ``democratic'' decays of light 
$2p$ emitters \cite{Egorova:2012,Golubkova:2016}.


\subsection{Sequential decay dynamics}
\label{sec:sequent}


\begin{figure}
\begin{center}
\includegraphics[width=0.48\textwidth]{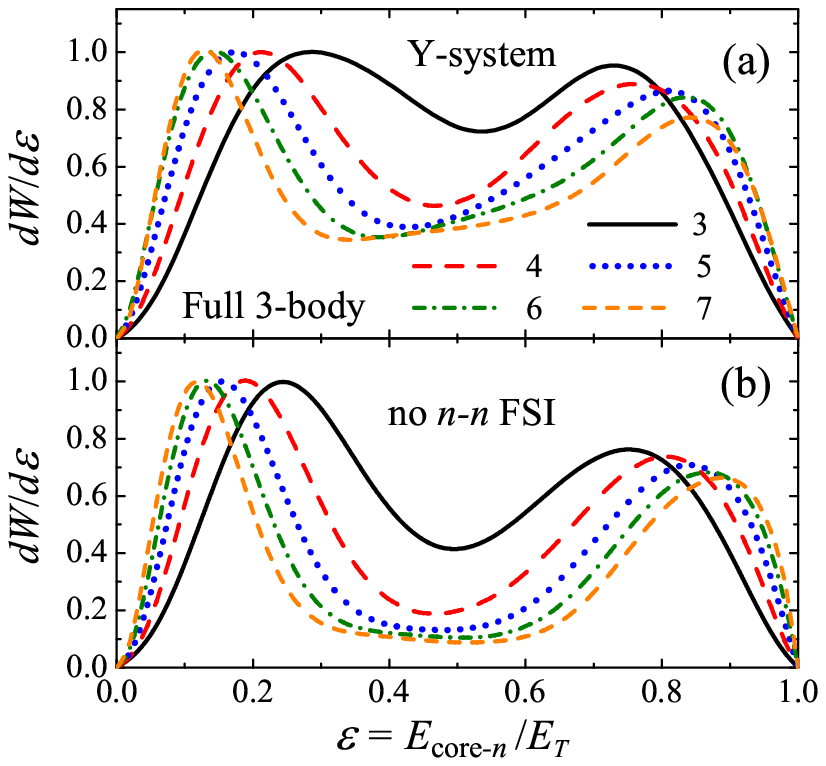}
\end{center}
\caption{Evolution of energy correlations for $^{6}$He SDM
transitions in the ``Y'' Jacobi system for $E_T = \{3,4,5,6,7 \}$ MeV. The
panels (a) and (b) correspond to full three-body calculation and ``no $n$-$n$
FSI'' case, respectively. All the left peaks in panel (a) have exactly the same 
absolute energy $\varepsilon E_T = 0.87$ MeV, which agrees very well with Fig.\ 
\ref{fig:5he} (b). }
\label{fig:eps-y-high}
\end{figure}

In the region (iii) with $2.5 \lesssim E_T $ MeV the decay regime is sequential:
the emission of nucleons proceed via population of the intermediate $p_{3/2}$
resonance in the $^{5}$He. This can be well seen in Fig.\ \ref{fig:eps-y-high}
as the two-hump structure of distributions for $E_T \gtrsim 2.5$ MeV. The
sequential decay peak drifts to lower $\varepsilon$ with $E_T$ increase.
However, it can be found from distributions of Fig.\ \ref{fig:eps-y-high} that
the peak with lower $\varepsilon$ value always takes place at the same energy
$E_r=0.84-0.86$ MeV. Where this energy $E_r$ is coming from?


\subsubsection{Sequential peak energy}

\begin{figure}
\begin{center}
\includegraphics[width=0.48\textwidth]{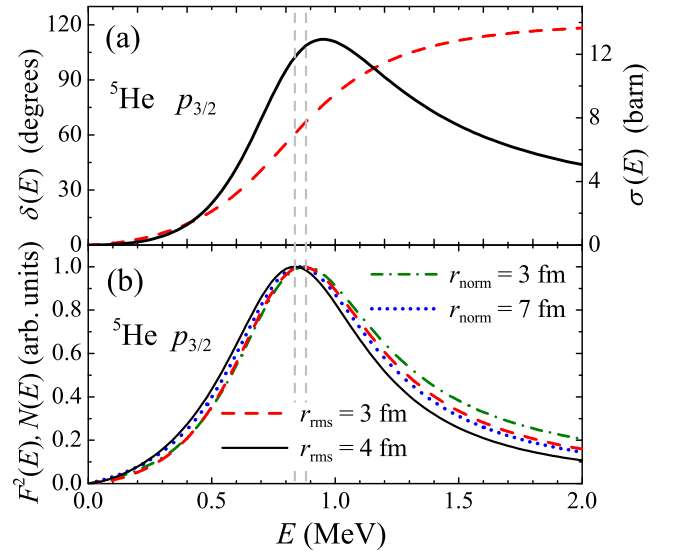}
\end{center}
\caption{Characteristics of the $^{5}$He $p_{3/2}$ resonance. Panel (a) shows
the phase shift and elastic cross section. Panel (b) shows internal
normalization Eq.\ (\ref{eq:int-norm}) with different $r_{\text{norm}}$ and 
continuum formfactors Eq.\ (\ref{eq:int-ff}) with different rms radii 
$r_{\text{rms}}$. Vertical dashed lines visualize positions of the 
highest-energy (0.83 MeV) and lowest energy (0.87 MeV) peaks in the panel (b).}
\label{fig:5he}
\end{figure}

The information on the $^{5}$He $p_{3/2}$ resonance, governing the properties of
the sequential decay, is given in Fig.\ \ref{fig:5he} (a). The standard
description of the resonance is represented by the elastic phase shift and the
corresponding elastic cross section. The elastic cross section for the potential
used in our calculations has the peak value at $E_r=0.95$ MeV. However, this
resonance is quite broad and we may question another continuum responses. Fig.\
\ref{fig:5he} (b) shows also the internal normalization
\begin{equation}
N_l(E) = \int _0^{r_{\text{norm}}} dr \, |f_l(kr)|^2 \,,
\label{eq:int-norm}
\end{equation}
where $r_{\text{norm}}$ is the size of the normalization region, and continuum 
formfactor
\begin{equation}
F_l(E) = \int _0^{\infty} dr \, f_l(kr) \phi(r) \,.
\label{eq:int-ff}
\end{equation}
Function $f_l(kr)$ is two-body scattering WF normalized as $\sin \sin(kr + l
\pi/2 + \delta_l)$ and the ``source'' WF $\phi(r)$ taken in a simple analytical
form (so-called Hulten Ansatz)
\begin{equation}
\phi(r) = \sqrt{\frac{2(r_{01}+r_{02})}{(r_{01}-r_{02})^2}} \, (\exp[-r/r_{01}]-
\exp[-r/r_{02}]) \,.
\label{eq:sour-ff}
\end{equation}
Here we use $r_{01}=0.5$ fm and vary $r_{02}$ to get different rms radii
for $\phi(r)$. It can be seen in Fig.\ \ref{fig:5he} (b) that the energies of
the peak both for internal normalization and for formfactors are considerably
different from the peak energy of elastic cross section. The $^{5}$He peak
energy $E_r=0.84-0.87$ MeV inferred from Fig.\ \ref{fig:5he} (b) is very stable
--- it has a small variation  when changing the parameters in a broad range of
``reasonable'' values. This range exactly corresponds to the stable peak
energies observed for different $E_T$ in Fig.\ \ref{fig:eps-y-high}.

\subsubsection{Convergence of momentum distributions}
\label{sec:mom-dis-conv}


Basing on our results which we have obtained from the studies of two-nucleon
emission and two-proton radioactivity we can conclude that for the
\emph{energies} of the resonant states the convergence is fastest, for
\emph{width} of this states it is slower, and for \emph{momentum correlations}
the convergence is the slowest \cite{Grigorenko:2007}.

\begin{figure}
\begin{center}
\includegraphics[width=0.48\textwidth]{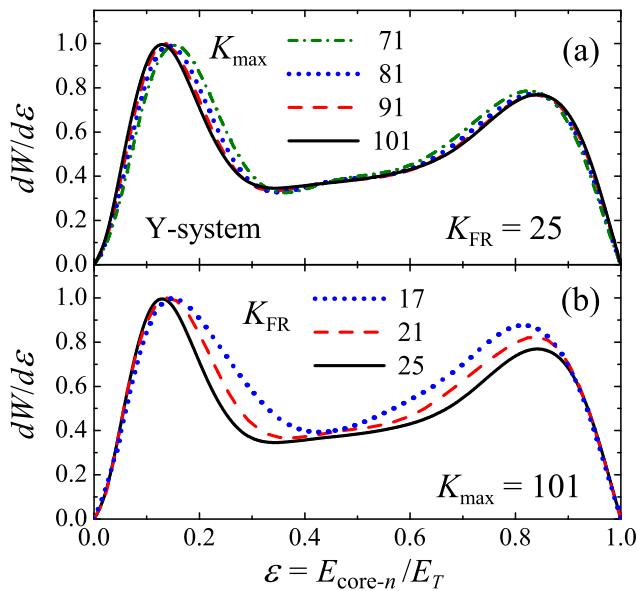}
\end{center}
\caption{Convergence of the energy distributions in the ``Y'' coordinate system
for $E_T=7$ MeV. (a) $K_{\max} $ convergence for fixed  $K_{FR} =25$. (b)
$K_{FR} $ convergence for fixed  $K_{\max} = 101$. }
\label{fig:eps-y-converg}
\end{figure}

We can find in Figs.\ \ref{fig:conv-low-k} and \ref{fig:conv-high-k} that the
convergence of strength function is very good for $K_{\max}>50$ in a broad
energy domain. If we look at the high energy part of SF with $E_T>5$ MeV, the
result is well converged already at very small basis sizes $K_{\max}\sim 13-15$.
In contrast, the convergence of the momentum distributions for the high-energy
part of the E1 strength function is found to be most challenging issue: here we
need to describe the long-range formation of the peaks in the distribution for
sequential decay mode via the $p_{3/2}$ resonant g.s.\ of $^{5}$He. Extremely
large basis sizes are required for that. It can be seen in Fig.\
\ref{fig:eps-y-converg} that the convergence is reasonably good, but not
quite achieved yet. This figure illustrates $E_T=7$ MeV case; for $E_T=3-5$
MeV the convergence is much better and it is perfect for the lower energies.

\subsubsection{``Anti-dineutron'' correlation}
\label{sec:anti}

\begin{figure}
\begin{center}
\includegraphics[width=0.48\textwidth]{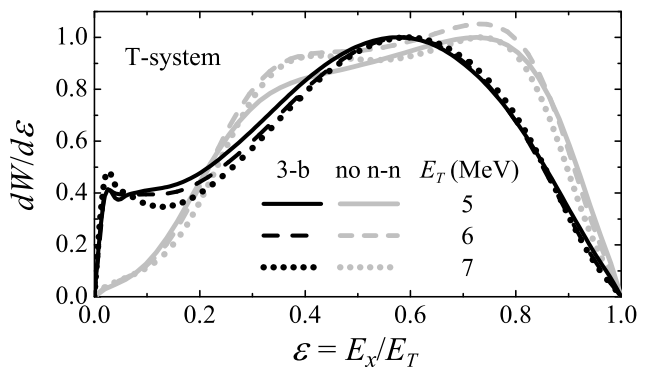}
\end{center}
\caption{Evolution of energy correlations for $^{6}$He SDM
transitions in the ``T'' Jacobi system for $E_T = \{5,6,7 \}$ MeV for the full
three-body calculation (black curves) and ``no $n$-$n$ FSI'' case (gray
curves).}
\label{fig:eps-t-high}
\end{figure}

The energy correlations in the Jacobi ``T'' system at energies $E_T \gtrsim 2.5$
 MeV are shown in Fig.\ \ref{fig:eps-t-high}. We can see that in contrast to the
``dineutron'' peak in the energy distribution at low $E_T$, a peculiar repulsive
anticorrelation takes place here between neutrons. In the calculations without
$n$-$n$ FSI there is strong suppression of probability for $\varepsilon \lesssim
0.2$. In full three-body calculation the $n$-$n$ FSI ``try'' to compensate this
effective repulsion somehow. The energy distribution even has a sharp increase
at $\varepsilon \rightarrow 0$. However, the intensity of the $n$-$n$ FSI is not
sufficient to overcome the overall repulsive trend: the probability for
$\varepsilon \lesssim 0.2-0.3$ is still seriously suppressed.

\subsection{Transitional decay dynamics}


In the region (ii) with $1 \lesssim E_T \lesssim 2.5$ MeV the transition from 
true three-body to sequential emission dynamics is taking place. Estimates show 
that from the penetrability point of view the turnover to sequential emission 
regime (via the $p_{3/2}$ resonance in the $^{5}$He) should take place at $E_T 
\gtrsim 1.2-1.5$ MeV. However, at energies $E_T \sim 1.8$ MeV, the energies of 
the ``first'' and the ``second'' emitted neutrons with respect to the 
$\alpha$-core are nearly equal, and, thus, both these nucleons can well populate 
the $^{5}$He $p_{3/2}$ resonance via its broad ``wings''. For that reason  the 
sequential decay  can not be formed up to $E_T \sim 2.5$ MeV and the emission 
has complex three-body character. The major trends of transitional dynamics were 
discussed in the paper \cite{Golubkova:2016}. The forms of transition we face 
here looks analogous to transitional dynamics observed in the light $2p$ 
emitters, such as $^{6}$Be \cite{Egorova:2012} and $^{16}$Ne 
\cite{Brown:2014,Brown:2015} (so-called ``democratic decay'').

The transition from the three-body to sequential regime is characterized by a
rapid qualitative change of the correlation patterns, see Fig.\
\ref{fig:en-ang-distr}. This is well illustrated by energy distribution changes
in Figs.\ \ref{fig:eps-t-low} and \ref{fig:eps-y-low}. Two most important
effects are taking place in the transitional energy range.

\noindent (i) The ``dineutron'' correlation between emitted neutrons
(enhancement at low $\varepsilon$ values) typical for the low-energy $E_T$ range
is ``dissolved'', and replaced with ``anti-dineutron'' correlation (depression
at low $\varepsilon$ values), as discussed above in Sec.\ \ref{sec:anti}.

\noindent (ii) The sequential decay patterns for population of the
intermediate $p_{3/2}$ resonance in $^{5}$He are formed: we start from
distribution with one peak at $\varepsilon \sim 0.5$ and end with sequential
two-peak correlation pattern.


\section{What can we change?}
\label{sec:what}


Before we discuss the previous theoretical results, we ask ourselves a natural
question: how stable are our predictions for E1 SF. To understand it we
investigated (i) the impact of the $^{6}$He g.s.\ WF variation and (ii) the
impact of the three-body potential $\tilde{V}_3$ variation, see Eq.\
(\ref{eq:shred-e1}). The variations are not necessarily realistic: our aim is to
find out which variations of the E1 SF in $^{6}$He are, in principle,
attainable.

\begin{table*}[bth]
\caption{Properties of different versions of the $^{6}$He g.s.\ WFs. Energies
are in MeV. The radial characteristics show root mean square values; $\langle 
r_{\alpha} \rangle  \equiv \langle r_3 \rangle$ is the rms distance from 
$\alpha$ cluster to $^{6}$He c.m. The last two columns show the E1 NEW sum rule 
value for $E_T<3$ MeV and the total value in $\text{e}^2\text{fm}^2$ units. }
\begin{ruledtabular}
\begin{tabular}[c]{cccccccccc}
Calculation & $E_{\text{b}}$ & $ \langle \rho \rangle $ &  $ \langle r_{\alpha} 
\rangle$ & $ \langle r_{nn} \rangle$ & $\Delta E_{\text{Coul}}$  &
$r_{\text{mat}}$ & $r_{\text{ch}}$ & $S^{(3)}_{\text{NEW}}$ &
$S^{(\infty)}_{\text{NEW}}$ \\
\hline
GPT $n$-$n$                & 0.973 & 5.16  &  1.17  & 4.50 & 2.302 & 2.43 &
2.019 & 0.568 & 1.307 \\
GPT $n$-$n$, strong $V_3$  & 1.1   & 5.02  &  1.14  & 4.41 & 2.400 & 2.39 &
2.002 & 0.514 & 1.241 \\
GPT $n$-$n$, weak  $V_3$   & 0.85  & 5.24  &  1.19  & 4.57 & 2.251 & 2.46 &
2.031 & 0.630 & 1.352 \\
BJ $n$-$n$                 & 0.973 & 5.10  &  1.15  & 4.48 & 2.345 & 2.41 &
2.008 & 0.562 & 1.262 \\
Mod.\ strong BJ $n$-$n$    & 0.973 & 5.53  &  1.31  & 4.52 & 2.095 & 2.57 &
2.103 & 0.854 & 1.639 \\
Mod.\ weak BJ $n$-$n$      & 0.973 & 4.66  &  0.99  & 4.44 & 2.680 & 2.26 &
1.922 & 0.317 & 0.936 \\
As in Ref.\ \cite{Danilin:1991}  & 0.973 & 5.49  &  1.23  & 4.88 & 2.111 & 2.54
&
2.048 & 0.672 & 1.445 \\
%
%
Experiment                 & 0.973 &       &        &     & 2.344 & 2.30(7) 
\cite{Egelhof:2001} & 2.068(11) \cite{Mueller:2007} & 0.45(12) 
\cite{Aumann:1999}     &      \\
                           &       &       &        &     &       &
                           2.48(3) \cite{Ozawa:2001}  &     &       &     \\
\end{tabular}
\end{ruledtabular}
\label{tab:he6-wf-all}
\end{table*}

The basic information about different versions of the $^{6}$He g.s.\ WFs is
provided in the Table \ref{tab:he6-wf-all}. The matter radius of $^{6}$He is 
obtained as
\[
6 \, r^2_{\text{mat}}(^6\text{He}) = \langle \rho \rangle ^2 + 4
\,r^2_{\text{mat}}(^4\text{He}) \,.
\]
The predicted matter radius of $^{6}$He lies somewhere in between two values 
extracted from experiment, which disagree with each other and, moreover, are 
quite old. The most restrictive observables are the Coulomb displacement energy 
in $A=6$ isobar and the charge radius. The $\Delta E_{\text{Coul}}$ value is 
reproduced nicely by our main calculation. The charge radius of $^{6}$He, which 
in the cluster model is given by
\[
r^2_{\text{ch}}(^6\text{He}) =  r^2_{\text{ch}}(^4\text{He}) + \langle 
r_{\alpha} \rangle ^2 + r^2_{\text{ch}}(n)\,,
\]
is a bit underestimated. However, we can not improve agreement for this 
characteristic without coming to contradiction. If we somehow expand the system 
to get correct charge radius, the agreement for $\Delta E_{\text{Coul}}$
will be get worse. It should be also understood that calculation of this value
in the cluster model depends on a number of parameters, and not all of them are
confidently defined. We use the following ingredients: $r_{\text{ch}}(p) = 0.84$
fm, $r_{\text{ch}}(^4\text{He})= 1.681$ fm, $r^2_{\text{ch}}(n) = -0.1161$ fm, 
also leading to the rms matter radius of the core cluster 
$r_{\text{mat}}(^4\text{He})=1.495$ fm.

\begin{figure}
\begin{center}
\includegraphics[width=0.49\textwidth]{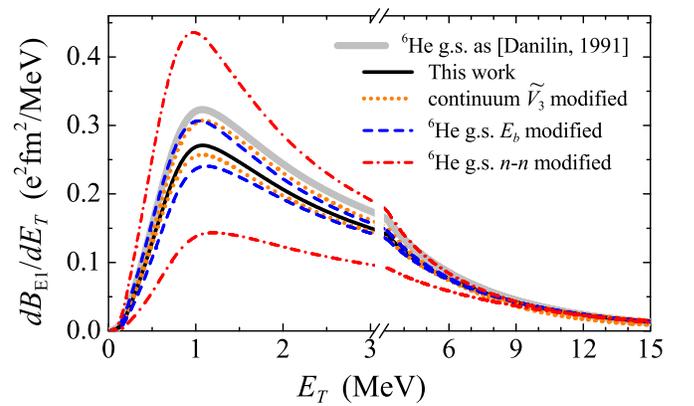}
\end{center}
\caption{What can we change by varying different aspects of the calculation. }
\label{fig:variat}
\end{figure}

The E1 SFs corresponding to different calculation options are collected in
Fig.\ \ref{fig:variat}.

\noindent (i) The ``old'' $^{6}$He g.s.\ WF from Ref.\ \cite{Danilin:1991}
produces the thick gray curve in Fig.\ \ref{fig:variat}.

\noindent (ii) A strong variation of the $\tilde{V}_3$ potential was performed,
see Eq.\ (\ref{eq:shred-e1}). The orange dotted curves correspond to
$V_3^{(0)}=32$ MeV (lower) and $V_3^{(0)}=-32$ MeV (upper). The scale of this
variation is unrealistically large. For example, for the $^{6}$He g.s.\
calculations the parameter $V_3^{(0)}=-13.5$ MeV is used to adjust the binding
energy to have exactly experimental value. We think that such variations of 
$\tilde{V}_3$ is much larger than any reasonable value: the many-body effects 
beyond the three-cluster approximation are expected to be smaller in $1^-$ 
continuum, compared to $0^+$ g.s.

\noindent (iii) The binding energy $E_b$ of $^{6}$He was varied by changing
$V_3$ potential, see Eq.\ (\ref{eq:shred-gs}). The blue dashed curves correspond
to $E_b=1.1$ MeV (lower) and $E_b=0.85$ MeV (upper).

\noindent (iv) The geometry of the $^{6}$He g.s.\ WF has been modified using
stronger and weaker $V_{nn}$ potentials. Red dash-dotted curves show the results
with $^{6}$He WF obtained with BJ potential [see Eq.\ (\ref{eq:bj})] with
$V_0=-21$ MeV (lower) and $V_0=-36$ MeV (upper), instead of the standard value
$V_0=-31$ MeV. The average angle between neutrons (as ``seen'' from the
$\alpha$-core) can be calculated as $60^{\circ}$, $66^{\circ}$, and $74^{\circ}$
for strong, normal, and weak $n$-$n$ potentials.

It can be understood from Fig.\ \ref{fig:variat} and Table \ref{tab:he6-wf-all}
that the result for E1 SF is quite stable. For variations (i) -- (iii) of
parameters far beyond realistic we get a modest change in the SF, which is
majorly just scaling within $\pm (10-15) \%$. To ``change'' the theoretical
prediction considerably [case (iv)], we need to change basic geometry of the
$^{6}$He g.s.\ WF. This can hardly be compatible with our common understanding
of structure, reactions, and observables for  $A = 6$ isobar.


\section{Discussion of previous theoretical results}

\begin{figure}
\begin{center}
\includegraphics[width=0.48\textwidth]{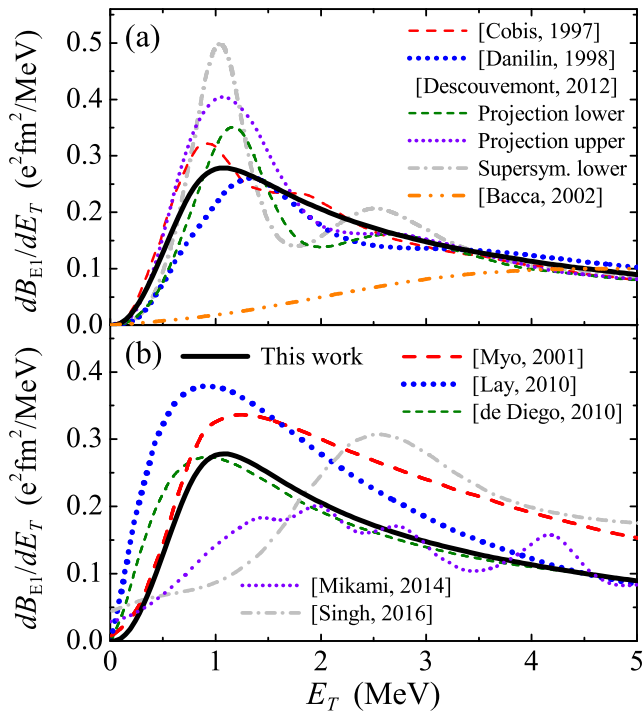}
\end{center}
\caption{Comparison of the results for the $^{6}$He E1 SF obtained in present
work with the previous calculations. Panel (a) shows the results of HH and
HH-based methods (Cobis, 1997: \cite{Cobis:1997}), (Danilin, 1998: 
\cite{Danilin:1998}), (Bacca, 2002 \cite{Bacca:2002}), (Descouvemont, 2012 
\cite{Descouvemont:2012}). Panel (b) shows the results of methods based on 
continuum discretization (Myo, 2001: \cite{Myo:2001}), (de Diego, 2010: 
\cite{deDiego:2010}), (Lay, 2010: \cite{Lay:2010}), (Mikami, 2014: 
\cite{Mikami:2014}), (Singh, 2016:  \cite{Singh:2016}).}
\label{fig:thcomp}
\end{figure}

Comparison of the results for the $^{6}$He E1 SF obtained in this work with the
previous calculations is given in Fig.\ \ref{fig:thcomp}.

Calculations \cite{Danilin:1998} were performed by members of our
collaboration in a very similar formalism, but with numerical limitations
natural to a situation twenty years ago. The ``wavy'' profile of the SF is
analogous to the results obtained in present work with limited
basis $K_{\max}=21$, see Fig.\ \ref{fig:conv-low-k}. The strength
function do not match exactly ours, obtained within the same limitations,
because the Pauli principle treatment in the core-$n$ channel is different in
\cite{Danilin:1998} (so-called ``Pauli projection'' method). Also the $^{6}$He
g.s.\ WF used in \cite{Danilin:1998} is somewhat different from ours.

The calculations of Refs.\ \cite{Cobis:1997} show the same ``wavy'' behavior
which, as we demonstrate in this work, is a symptom of insufficient convergence.
The computational methods of \cite{Cobis:1997} and of present work are
different, but both rely on hyperspherical expansion of WF. Therefore, we may
still expect some analogy in convergence trends. The behavior of the strength
function in Refs.\ \cite{Cobis:1997} on the left slope of the peak ($E_T \sim
0.2-0.8$ MeV) is very close to the behavior of our strength function. Basing on
our experience, we do not expect that this aspect of the SF \cite{Cobis:1997}
changes noticeably in the case of the complete convergence.

The HH calculations of Ref.\ \cite{Descouvemont:2012} again show the ``wavy''
behavior discussed above. Several calculations were presented in this work,
divided in two groups by treatment of the Pauli principle in the $\alpha$-core
channel: (i) ``Pauli projection'' and (ii) ``supersymmetric transformation''
tecniques. We have selected the upper and lower results from group (i) and the
lower from group (ii) --- the other results from this group look a bit
unrealistic.

\begin{figure}
\begin{center}
\includegraphics[width=0.48\textwidth]{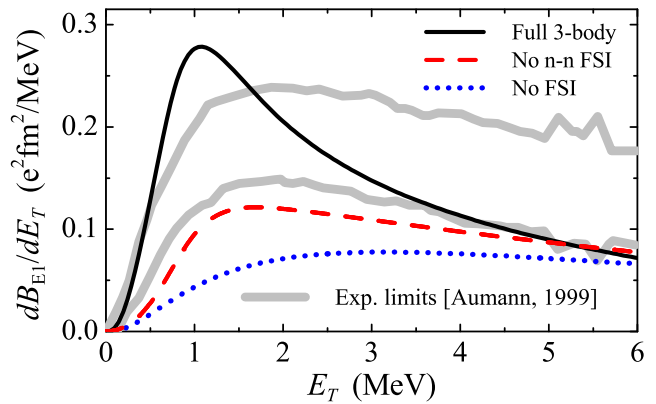}
\end{center}
\caption{Comparison of the results for the $^{6}$He E1 SF obtained in present
work with experimental data \cite{Aumann:1999}.}
\label{fig:expcomp}
\end{figure}

It can be seen in Fig.\ \ref{fig:thcomp} (a) that all the presented results from
\cite{Descouvemont:2012} (actually all the results of HH-based methods) are a
kind of oscillating around the ``mean value'' represented by our fully converged
calculations. On the other hand, it should be noted that all these methods give
qualitatively very similar SFs in the low-energy range, conforming the expected
\begin{equation}
\frac{dB_{\text{E1}}}{dE_T} \sim E^3_T \,,
\label{eq:sf-ass}
\end{equation}
behavior of the E1 SF for $^{6}$He. Having this correct low-energy asymptotics
is the natural feature of the HH method.

The results of  Ref.\ \cite{Bacca:2002} were obtained in the HH-based method as 
well. However, this is 6-body aproach treating photodissociation in the special 
framework (Lorentz integral transformation). The three-body cluster 
$\alpha$+$n$+$n$ threshold is not explicitely present in this approach (only the 
6-body threshold) and the low-energy behavior in this channel can be provided 
only by the basis convergence. However the maximum basis size achieved in this 
work is only $K_{\max}=11$, which is far not sufficient even in the three-body 
case, which is technically much easier. As a result the effective strength 
function (recalculated very approximately from the cross section given in 
\cite{Bacca:2002}) does not demonstrate ``soft dipole'' low-energy enhancement 
which takes place in some form in all other approaches.

The theoretical methods collected in Fig.\ \ref{fig:thcomp} (b) are based on
different forms of the continuum discretization.

The calculations of \cite{Singh:2016} (this is essentially three-body model) 
give a peak in the E1 SF at about 2.5 MeV, which is too different from the 
``mainstream'' value of $1.0-1.1$ MeV.

\begin{figure}[t]
\begin{center}
\includegraphics[width=0.48\textwidth]{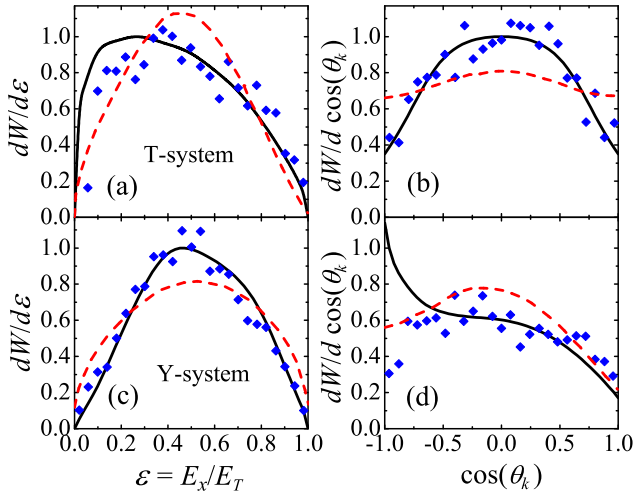}
\end{center}
\caption{The energy (a,c) and angular (b,d) distributions for products of
the E1 dissociation of $^{6}$He in ``T'' (a,b) and ``Y'' (c,d) Jacobi coordinate
systems obtained in the continuum energy region $E_T = 1-3$ MeV. The
experimental data from Ref.\ \cite{Chulkov:2005} are shown by diamonds. The
solid black curves correspond to the calculation results of present work. The
red dashed curves show the results of model Ref.\ \cite{Danilin:1998} as given
in \cite{Chulkov:2005}.}
\label{fig:expcomp-e13}
\end{figure}

The 6-body variational method calculations of \cite{Mikami:2014} effectively 
takes into account diffeent possible kinds of clusterization beside 
$\alpha$+$n$+$n$. This method provides low-energy concentration of the E1 
strength at about 2 MeV, which is higher than ``mainstream'' $1.0-1.1$ MeV value 
but can be seen as a nice result considering complexity of the approach. The 
low-energy spectrum below 3 MeV is based on 6-8 discrete states and thus its 
specific profile is stronfly sensitive to the smoothing procedure.

The strength function obtained in Ref.\ \cite{Myo:2001} has a peak at higher 
energy ($E_T = 1.25$ MeV) than in present work and in the papers Ref.\ 
\cite{Cobis:1997,deDiego:2010} ($E_T \sim 0.9-1.1$ MeV). This could be a result 
of incomplete accounting of the $n$-$n$ FSI. The peak energy in \cite{Myo:2001} 
is close to 1.4 MeV (value obtained in calculations without $n$-$n$ FSI), see 
Fig.\ \ref{fig:thcomp} (b). Another worrying issue is the high-energy behavior 
of the SF obtained in the work \cite{Myo:2001}. The SF is shown up to 6 MeV 
only, but if we smoothly extrapolate it to higher energies, we can infer that 
the E1 NEW sum rule value for this SF is around 1.8 e$^2$fm$^2$. This value 
corresponds to $r_{\alpha}= 1.37$ fm rms distance of $\alpha$ cluster from the 
center of mass in the $^{6}$He g.s. This is considerably larger than 
$r_{\alpha}= 1.17$ fm for $^{6}$He WF used in the present work (similar radial 
properties of $^{6}$He WF were used also in the calculations 
\cite{Cobis:1997,deDiego:2010}). The rms matter radius of the $^{6}$He is 
$r_{\text{mat}}= 2.43$ fm in our work (based on the 1.495 fm rms matter radius 
of $\alpha$-cluster) and the corresponding $r_{\text{mat}}= 2.46$ fm can be 
found in \cite{Myo:2001}. The difference here is not that large. Thus we have to 
presume very different geometry of $^{6}$He in the work \cite{Myo:2001}: the rms 
distance between two neutrons recalculated from $r_{\alpha}$ and 
$r_{\text{mat}}$ is $r_{nn}= 3.11$ fm. This is drastically smaller than the 
value $r_{nn}= 4.49$ fm used in present work. In general, the values $r_{nn}> 
4.3$ fm are typically found in all other model calculations of $^{6}$He.

The $^{6}$He E1 SF of \cite{deDiego:2010,Lay:2010} noticeably differs  from our
SF and from SF in Refs.\ \cite{Cobis:1997,Descouvemont:2012}. Namely, the
low-energy behavior of the SF in \cite{deDiego:2010,Lay:2010} is strongly
enhanced compared to the HH-based works. Such a behavior is very difficult to
reproduce in realistic calculations. This is an important issue since
astrophysical capture rates are very sensitive to the low-energy behavior of the
SF in a broad range of temperatures of interest. Therefore, we will return to
this problem in the forthcoming publication Ref.\ \cite{Grigorenko:2020b}. It
seems that all the methods collected in Fig.\ \ref{fig:thcomp} (b) have problems
with correct treatment of the low-energy part of the E1 SF in $^{6}$He. All
these SFs tend at $E_T \rightarrow 0$ either to constant, or to something
visually different from the expected behavior of Eq.\ (\ref{eq:sf-ass}), see
Fig.\ \ref{fig:thcomp} (a).

The correlation aspect of the SDM in $^{6}$He has been fragmentarily discussed
in papers \cite{Danilin:1998,Myo:2014}. The results of \cite{Danilin:1998} we
discuss below when comparing with experimental data. In paper \cite{Myo:2014}
the inclusive $E_{n\text{-}n}$ and $E_{\text{core-}n}$ distributions were
constructed for the Coulomb breakup reaction. No comparison of this information
with our results is possible.


\section{Comparison with experimental data}


\begin{figure}[t]
\begin{center}
\includegraphics[width=0.48\textwidth]{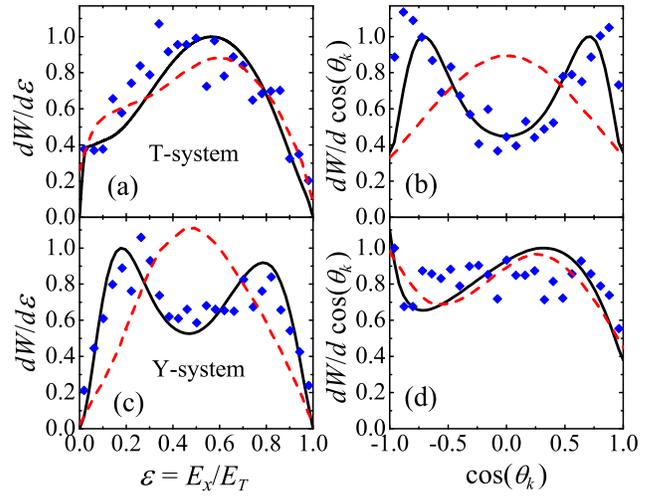}
\end{center}
\caption{The same as in Fig.\ \ref{fig:expcomp-e13}, but for the continuum
energy region $E_T = 3-6$ MeV.}
\label{fig:expcomp-e36}
\end{figure}

The results of the $^{6}$He E1 dissociation calculations are compared with
available experimental data in Fig.\ \ref{fig:expcomp} for SF and in Figs.\
\ref{fig:expcomp-e13}, \ref{fig:expcomp-e36} for fragment correlations.

Our SF, as well as other predictions
\cite{Cobis:1997,Danilin:1998,Myo:2001,deDiego:2010}, are consistent with each
other and with experimental data \cite{Aumann:1999} for $E_T> 2$ MeV. However,
for lower energies there is strong disagreement between experiment and all
the calculations. There are certain differences in details which we discussed
above, but all the theoretical calculations in Fig.\ \ref{fig:expcomp} predict a
pronounced peak of E1 SF in $^{6}$He  at $E_T \sim 0.9-1.25$ MeV with
peak values in the range $\sim 0.27-0.33$ e$^2$fm$^2$/MeV. This feature of all
the existing calculations disagrees with data far beyond the experimental
uncertainty declared in \cite{Aumann:1999}.

Let us take a look at the correlation patterns observed for the E1 dissociation
of $^{6}$He in the experiment Ref.\ \cite{Aumann:1999} and later published in
Ref.\ \cite{Chulkov:2005}. It should be understood that the theoretical
distributions are shown together with experimental data ``as is'', without any
accounting for the experimental bias, so this comparison can not be truly
quantitative. Nevertheless, in Figs.\ \ref{fig:expcomp-e13} and
\ref{fig:expcomp-e36} we may see that the agreement of the theoretical
correlations with experimental data significantly  improves, compared to the
predictions of Ref.\ \cite{Danilin:1998}. For example, there was a strong
qualitative disagreement between theory of \cite{Danilin:1998} and experiment in
Fig.\ \ref{fig:expcomp-e13} (b,c), which is ``cured'' in our modern
calculations.

There is, however, a considerable disagreement with the experimental data in
correlations, which we should emphasize. The data is much smaller than theory in
the energy distribution Fig.\ \ref{fig:expcomp-e13} (a) for $\varepsilon < 0.3$
and in the angular distribution Fig.\ \ref{fig:expcomp-e13} (d) for $\cos
(\theta_k) < -0.6$. Both these ranges correspond to the same physical situation
of low momentum between two neutrons. One may see in Fig.\ \ref{fig:eps-t-low}
(a) that the energy distribution around $E_T\sim 1$ MeV has an pronounced
low-energy $n$-$n$ peak (at higher $E_T$ energies the low-energy $n$-$n$
correlation is supressed, see also Fig.\ \ref{fig:eps-t-high}). If we assume
that the efficiency of the registration of the low-energy two-neutron events was
\emph{underestimated} in the data treatment of \cite{Aumann:1999}, then both of
these disagreements in correlations and the absence of the $E_T \sim 1 $ MeV
peak in experimental reconstructed strength function in Fig.\ \ref{fig:expcomp}
get explanation. To clarify this issue new high precision experiments dedicated
to SDM in $^{6}$He are necessary.


\section{Conclusions}


Accurate calculations of the E1 strength function (or soft dipole mode) for
$^{6}$He are presented in this work. The results of these calculations
significantly improved  the older results of the same collaboration  Ref.\
\cite{Danilin:1998}. Both the E1 strength function and three-body decay
correlation pattern are found to be fully converged for $E_T > 0.1$ MeV. Fully
converged results of this work allow to understand strange ``wavy'' behavior of
the E1 SF predicted in theoretical papers
\cite{Danilin:1993,Cobis:1997,Descouvemont:2012} and corresponding strong
disagreement among them. For the first time we are able to get insight for the
decay dynamics for the soft dipole excitations. The transition from ``true''
three-body decay dynamics to sequential decay mechanism is taking place in the
energy range $E_T= 1.0-2.5$ MeV which is reflected in the evolution of the
correlation patterns.

It was demonstrated that the SDM results, obtained in this work, have important
advantages compared to the other available theoretical calculations. However,
the interpretation of the soft dipole excitation is a ``fragile'' issue, very
sensitive to details of the models. Therefore the theoretical results have to be
confirmed by the experimental data. Unfortunately, there is no agreement between
E1 SF for $^{6}$He obtained in the calculations and the E1 strength function
extracted from the $^{6}$He Coulomb dissociation cross section on the heavy
target for the low-energy range $E_T<2$ MeV. This is true not only for our
calculations, but for all the theoretical results available so far. Such a
situation is unsatisfactory, since the radiative capture rates in nuclear
astrophysics can be obtained only by the extrapolation based on the Coulomb
dissociation cross section data. This is exactly the low-energy range, where the
quality of the data are crucial for extrapolation to energies of astrophysical
interest.

Intensive $^{6}$He beams are easily accessible at the modern RIB facilities, so
it would be expected that the measurements with $^{6}$He should become a
benchmark case for all the studies of this kind. So, highly precise measurements
of the E1 SF in $^{6}$He  with modern techniques, as well as scrupulous
comparison with theoretical calculations, taking the experimental bias into
account, are very desirable.



\paragraph*{Acknowledgments.} ---
LVG and NBS were supported in part by the Russian Science Foundation grant No.\
17-12-01367.


\bibliographystyle{apsrev4-1}
\bibliography{d:/latex/all}


\end{document}